\documentclass[twocolumn,english,aps,prb,showpacs,superscriptaddress,amssymb,amsfonts]{revtex4}
\usepackage{amsmath}
\usepackage{graphicx}
\usepackage{amssymb}
\usepackage{color}
\usepackage{tabularx}
\makeatletter
\newcommand{\be}{\begin{equation}}
\newcommand{\ee}{\end{equation}}
\newcommand{\bea}{\begin{eqnarray}}
\newcommand{\eea}{\end{eqnarray}}

\makeatother

\makeatother

\usepackage{babel}

\makeatother

\usepackage{babel}

\begin{document}

\title{Entanglement Entropy as a Portal to the Physics of Quantum Spin Liquids}

\author{Tarun Grover}
\affiliation{Department of Physics, University of California,
Berkeley, CA 94720,
USA}
\affiliation{Kavli Institute for Theoretical Physics, University of California, Santa Barbara, CA 93106, USA}
\author{Yi Zhang}
\affiliation{Department of Physics, University of California,
Berkeley, CA 94720,
USA}
\author{Ashvin Vishwanath}
\affiliation{Department of Physics, University of California,
Berkeley, CA 94720,
USA}

\begin{abstract}
Quantum Spin Liquids (QSLs) are phases of interacting spins that do not order even at the absolute zero temperature, making it impossible to characterize them by a local order parameter. In this article, we review the unique view provided by the quantum entanglement on  QSLs. We illustrate the crucial role of Topological Entanglement Entropy in diagnosing the non-local  order in QSLs, using specific examples  such as the Chiral Spin Liquid. We also demonstrate the detection of anyonic quasiparticles and their braiding statistics using quantum entanglement. In the context of gapless QSLs, we discuss the detection of emergent fermionic spinons in a bosonic wavefunction, by studying the size dependence of entanglement entropy.

\end{abstract}

\maketitle
\tableofcontents

\section{Introduction} \label{sec:intro}

Quantum Spin Liquids (QSLs) are phases of strongly interacting spins that do not order even at absolute zero 
temperature and therefore, are not characterized by a Landau order
parameter \cite{balentsreview}. They are associated with remarkable phenomena such as
fractional quantum numbers \cite{anderson1973, anderson1987, baskaran1987, read1991}, transmutation of
statistics (eg. fermions appearing in a purely bosonic
model)\cite{kivelson1987, read1989}, and enabling otherwise
impossible quantum phase transitions\cite{senthil2004}, to name a
few. Spin liquids may be gapless or gapped. While current
experimental candidates for spin liquids appear to have gapless
excitations \cite{exp},  gapped spin liquids are indicated in
numerical studies on the Kagome \cite{yan2010, hongchen2012}, square \cite{Balents, Wen} and honeycomb lattice
\cite{meng2010}. A distinctive property of gapped spin liquids is that they are characterized by
topological order - i.e. ground state degeneracy that depends on the
topology of the underlying space\cite{wen2004, hastings2004}. The gapless spin liquids are relatively difficult to characterize and their low-energy
description often involves strongly interacting matter-gauge theories \cite{mattergauge}.

\begin{figure}
\begin{centering}
\includegraphics[scale=0.35]{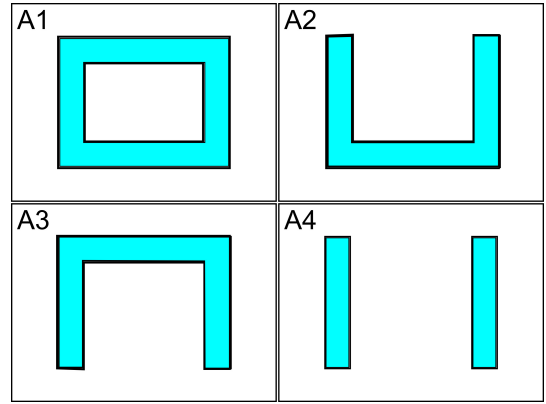}
\par\end{centering}
\caption{Following Ref.~\onlinecite{levin2006}, consider four regions $A_1$-$A_4$ (shaded cyan). The thickness as well as the linear size of all regions $A_i$ is much bigger than the correlation length $\xi$ of the system. The combination
$(S_1 - S_2) - (S_3 - S_4)$ equals zero for short-range entangled systems. To see this, we note that $S_1 - S_2$ and $S_3 - S_4$ correspond to the change in entanglement entropy associated with closing the region $A_2$ and $A_4$, respectively, at the top. For a wavefunction with short-range entanglement, the difference $(S_1- S_2)-(S_3 -S_4)$ should vanish as $e^{-R/\xi}$ where $R$ is  approximately the linear size of regions $A_i$. Therefore, a non-zero $\gamma$ implies 
that the system is long-range entangled.} \label{gamma}
\end{figure}

In the theoretical search for QSLs, two qualitatively different kinds of questions can be asked:

\begin{itemize}
 \item Given a particular Hamiltonian $H$, does it admit a QSL as its ground-state? If yes, can one further characterize the QSL?

\item Given a ground state wavefunction(s), can one tell if it corresponds to a QSL? If yes, can one further characterize the QSL?
\end{itemize}

These two questions are related: if a numerical scheme can provide one with access to approximate ground state wavefunction of a particular Hamiltonian $H$, then answering the
second question leads to an answer to the first one as well. The recent years have seen remarkable progress in tackling these questions. Of particular usefulness has been the application of ideas from quantum information science
to quantum many-body physics. Motivated by these developments, in this review we will focus on the second question above from an information theoretic point of view, and we will discuss some of the advances on the first question in the section \ref{sec:synop}.

The usefulness of entanglement entropy to understand the physics of
QSLs becomes apparent when one introduces the notion of
`short-range' and `long-range' entanglement. The heuristic picture
of a short-range entangled phase is that its caricature ground state
wavefunction can be written as a direct product state in the real
space, hence the nomenclature. The phases of matter that do not
satisfy this property are thereby called `long-range entangled'.
From this definition, gapless phases, such as a Fermi liquid or
superfluid, that exhibit algebraic decay of local operators are
always long-range entangled in the real space. Naively, one might
think that the converse would hold, that is, existence of
short-range correlations for local operators will imply short-range
correlation. It is precisely here that the connection between gapped
QSLs and ground state entanglement properties comes into play. To
wit, gapped QSLs are phases of matter that show long-range
entanglement in their ground states despite the fact that the
correlation functions of all local operators decay exponentially
with distance.

To make the connection between gapped QSLs and long-range entanglement precise, let us define some basic terminology that will also be useful throughout this review.
Given a normalized wavefunction, $|\phi\rangle$, and a partition of the system into subsystems $A$ and $B$, one can trace out the subsystem B to give a
 density matrix on A: $\rho_A = {\rm Tr}_B |\phi\rangle\langle\phi|$.
The Renyi entropies are defined by:
\begin{equation}
S_n = \frac1{1-n}\log({\rm Tr}\rho_A^n) \label{eq:defee}
\end{equation}
It is common to pay special attention to the von Neumann entropy, $S_1= -{\rm Tr}[\rho_A \log \rho_A]$ (obtained by taking the limit $n \rightarrow 1$ in Eqn.\ref{eq:defee}). However, the Renyi entropies are
also equally good measures of entanglement, and are often easier to compute.

In seminal papers by Kitaev, Preskill \cite{kitaev2006} and Levin,
Wen \cite{levin2006}, it was shown that the entanglement entropy of
a topologically ordered phase, such as a gapped QSL, as a function of the linear size of
disc-shaped region $A$ of linear size $l$, receives a universal subleading contribution:

\be
S(L) = \frac{l}{a} - b_0 \gamma + O(1/l) \label{eq:teeintro}
\ee

\noindent where $b_0$ is the number of connected components of the boundary of region $A$ and the constant $\gamma$ is a universal property of the phase of matter that is non-zero if and only if the phase is topologically ordered.
One may wonder why does a non-zero $\gamma$ implies
long-range entanglement? One way to understand this \cite{levin2006} is to  consider the combination $S_0 = (S_1 - S_2) - (S_3-S_4)$ for the geometry shown in Fig.\ref{gamma}. If the phase was in fact short-range entangled,
then $S_0$ will vanish identically. From Eqn.\ref{eq:teeintro}, $S_0 = - 2\gamma$ and therefore, a non-zero $\gamma$ implies long-range entanglement. A different way to derive the same conclusion is
to define a quantity $S_{local}$ such
that it is obtained by locally patching the contributions to the entanglement entropy
from the boundary $\partial A$ of region $A$ \cite{grov2011}. If the phase is short-range entangled, then $S_{local}$ is the entire $S$. As shown in Ref.~\onlinecite{grov2011}, on general grounds, $S_{local}$ cannot contain a constant term,
and therefore, $\gamma \neq 0$ implies long-range entanglement.

Gapless spin-liquids pose a different set of questions. The gapless excitations make it difficult to assign them a simple topological number such as $\gamma$ above. One exception is 
gapless spin-liquids where the emergent gauge degrees of freedom are gapped and are coupled to gapless matter. In this case, there are known contributions to the entanglement entropy that
can characterize the nature of QSL \cite{senthilswingle, pufu}. In this review, we will focus on cases where the 
emergent gauge fields are instead gapless. As an example, consider a two-dimensional QSL where the gapless matter fields are neutral fermions which form a Fermi surface, and are coupled to a gapless 
gauge field. As we discuss in Sec.\ref{sec:gapless}, empirically one finds that their entanglement entropy scales as $S \sim l\log l$, akin to gapless fermions in two dimensions \cite{gioev2006, wolf}, thus suggesting the presence of emergent fermions in a purely bosonic state.

\section{Entanglement Entropy and Many Body Physics} \label{sec:manybody}

One of the most interesting properties of EE is that $S(A) =
S(\overline{A})$ where $\overline{A}$ is the complement of region
$A$. This suggests that EE is a property of the boundary $\partial
A$ of region $A$ since $\partial A = \partial \overline{A}$, which
further suggests that for a short-ranged correlated system, EE
scales linearly with the the boundary $\partial A$: $S(l) \sim l^{d-1}$ 
where $l$ is the linear size of region $A$ \cite{bombelli, srednicki}.
Such a scaling of EE goes by the nomenclature of ``Area Law'' or
``Boundary Law''. Indeed, for one-dimensional systems with a gap,
this statement can be proved rigorously \cite{hastingsproof} and it has also been
observed numerically for various gapped two-dimensional systems \cite{eisert2010}.
Furthermore, even when the correlations of local operators are
long-ranged, for example, in a free fermionic system with a Fermi
surface, the Boundary Law is violated by only by a multiplicative
logarithm, that is, $S(l) \sim l^{d-1} \log (l)$ \cite{gioev2006, wolf}, while for a
conformal field theory (CFT), where the low-lying modes exist at
only discrete point(s) in the momentum space, the Boundary Law is
not violated at all in $d >1$ \cite{ryu2006}.

To give a flavor of the information theoretic approach to many-body physics, let us consider one of the earliest such applications that
concerns the relation between the central charge $c$ of a one-dimensional CFT, and
the coefficient of the logarithmic term in the bipartite entanglement entropy $S$ for the corresponding CFT \cite{cftwork}. The result is

\be
S(l) = \frac{c}{6} \log(\frac{l}{a}) \label{eq:onedcft}
\ee

\noindent where $l$ is the size of the subsystem $A$ and $a$ is a microscopic cutoff. The main observation to gather from this equation is that just given the ground state wavefunction of a CFT,
its central charge can be extracted by calculating the entanglement entropy. Note that rescaling $l \rightarrow l/\alpha$ where $\alpha$ is a non-universal number does not change the
coefficient of the logarithm. This makes physical sense, since $c$ is a property of the low-energy theory and should be insensitive to the details of how the theory
is regularized at the level of the lattice.

An extension of this result to a two dimensional CFT is \cite{ryu2006}:

\be
S(l) = \frac{l}{a}- \Gamma \label{eq:twodcft}
\ee

\noindent  for a region $A$ with linear size $l$ and smooth boundaries. Here
$\Gamma$ is a constant that depends only on the shape of region $A$.
Rescaling $l \rightarrow l/\alpha$ does not change $\Gamma$ akin to
the case of one-dimensional CFT, albeit in this case $\Gamma$ will generically be shape dependent. As we will
see, this result will be useful to characterize gapless QSLs whose
low-energy theory is a CFT (Sec.\ref{sec:gapless}).

The common feature of the above two examples is that EE as a function of system size $l$ contains terms that are universal properties of the low-energy description of the system. Returning to
our topic of interest, namely QSLs, one of the most interesting result \cite{kitaev2006, levin2006} is that for gapped systems in two dimensions, the subleading term in EE
is universal and contains information regarding the presence or absence of topological order, and also partially characterizes the topological order:

\be
S(l) = \frac{l}{a} - \gamma + O(1/l) \label{eq:teedef}
\ee

The subleading constant $\gamma$ is called ``Topological Entanglement Entropy'' (TEE) and as already mentioned in the introduction, is a universal property of the phase of the matter,
even though one happens to have has access to only one of
the ground states in the whole phase diagram. $\gamma$ is non-zero if and only if the system has topological order. Furthermore,
$\gamma = \log(D)$ where $D = \sqrt{\sum_i d^2_i}$ is the so-called total quantum dimension associated with the phase of matter whose quasiparticles carry a label $d_i$, the individual quantum
dimension for the $i$'th quasiparticle. As a concrete example, for Kitaev's Toric code model \cite{kitaev2003}, there are four quasiparticles, the identity particle $\mathbf{1}$, the electric charge $\mathbf{e}$,
the magnetic charge $\mathbf{m}$, and the dyon $\mathbf{em}$. The quantum dimension is one for each of these quasiparticles and hence the total quantum dimension $D = \sqrt{4} = 2$.
We note that the the universal contribution to EE was first observed in the context of discrete gauge theories in Ref.~\onlinecite{hamma2005}. 

The Eqn.\ref{eq:teedef} serves as a definitive test to see if a particular ground state corresponds to topological ordered state or not. If the answer is indeed in affirmative and
if the state also possesses a global $SU(2)$ spin-rotation symmetry, then it is a gapped QSL. Having established that the wavefunction corresponds to QSL, it is imperative to further characterize
 the state further since the total quantum dimension $D$ only a  partial characterization of topological order. For example, for $D = 2$, there exist two distinct QSLs, the $Z_2$ QSL, which has the same topological order
as the Kitaev's Toric code \cite{kitaev2003}, and the doubled semionic theory \cite{levin2005}, which can be thought of as a composite of Laughlin $\nu = 1/2$ quantum Hall state and its time-reversed partner.
To achieve a detailed characterization of topological order, TEE again turns out to be a very useful concept. As we will show in Sec.\ref{sec:stat}, TEE can be used to extract not only the
individual quantum dimensions $d_i$ for the quasiparticles, but there braiding statistics as well.

An elegant different approach to a more complete identification of topological order is through the study of the entanglement spectrum \cite{HaldaneLi}. This method is extremely useful for  phases that have a gapless boundary such as quantum Hall and topological insulators/superconductors, though it may not be applicable for topological phases such as  the $Z_2$ spin liquid that do not have edge states.

\section{Variational Wavefunctions for Quantum Spin Liquids} \label{sec:varwfn}

A quantum spin liquid (QSL) can be viewed as a state where each spin
forms a singlet with a near neighbor, but the arrangement of
singlets fluctuates quantum mechanically so it is a liquid of
singlets \cite{anderson1973, balentsreview}. Theoretical models of this singlet liquid fall roughly
into two categories. In the first, the singlets  are represented as
microscopic variables as in quantum dimer and related
models\cite{rokhsar1988, moessner2001, senthil2002,
balents2002,misguich}, and are suggested by large N calculations
\cite{read1991}.  Topological order can then be established by a
variety of techniques including exact solutions. In contrast there
has been less progress establishing topological order in the second
category, which are SU(2) symmetric spin systems where valence bonds
are emergent degrees of freedom. More than two decades ago,
Anderson\cite{anderson1987} proposed constructing an SU(2) symmetric
spin liquid wavefunction by starting with a BCS state, derived from
the mean field Hamiltonian:

\be H = -\sum_{rr'} \{t_{rr'}f^{\dagger}_{\sigma,r} f_{\sigma,r'} +
\Delta_{r r'} f^{\dagger}_{\uparrow,r} f^{\dagger}_{\downarrow,r'}\}
+ h.c. \label{eq:meanfieldbcs} \ee

\noindent  and Gutzwiller projecting \cite{gutzwiller1963} it so that there is exactly one fermion
per site, hence a spin wavefunction. Variants of these are known to
be good variational ground states for local spin Hamiltonians (see
e.g. \cite{motrunich05, varstudy}) and serve as a compact description of most
experimental and $SU(2)$ symmetric liquids. Approximate analytical
treatments of projection, that include small fluctuations about the
above mean field state, indicate that at least two kinds of gapped
spin liquids can arise: chiral spin liquids\cite{kalmeyer} (CSL) and $Z_2$
\cite{	read1991,wen1991,senthil2000} spin liquids. The low energy theory of CSL corresponds to a Chern-Simons theory while in the $Z_2$ spin-liquid,
it corresponds to deconfined phase of the $Z_2$ gauge theory. However, given
the drastic nature of projection, it is unclear if the actual
wavefunctions obtained from this procedure are in the same phase as the ground state of a Chern-Simons theory/$Z_2$ gauge theory.
In fact, detecting deconfinement in a gauge theory in the presence of matter-field is considered notoriously hard \cite{fradkinshenker}. As we will see below, quantum entanglement
provides an elegant solution to this problem.

In contrast to gapped QSLs, the low-energy description of gapless QSLs consist of strongly interacting matter-gauge theories which are harder to analyze \cite{mattergauge}.
Remarkably, the above procedure of constructing mean-field spin-liquid states and using Gutzwiller projection to obtain wavefunctions can be applied to obtain gapless QSLs as well.
To do so, one sets the BCS gap $\Delta_{rr'}$ in Eqn.\ref{eq:meanfieldbcs}  to zero. This Gutzwiller projection technique is known to work well in one dimension where the projected Fermi sea spin wavefunction captures
 long distance properties of the Heisenberg chain, and is even known to be the exact ground state of the Haldane-Shastry \cite{HaldaneSastry} Hamiltonian. Though similar rigorous results are not available in two dimension,
a critical spin liquid, the spinon Fermi sea (SFS) state, has been invoked \cite{motrunich05, palee2005} to account for the intriguing phenomenology of aforementioned triangular lattice organic compounds \cite{yamashita2010}.
 In the SFS state, the spinons $f$ hop on the triangular lattice sites giving rise to a Fermi sea, while strongly interacting with an emergent `electromagnetic' $U(1)$ gauge field. The metal like specific heat and thermal
 conductivity seen in these materials is potentially an indication of the spinon Fermi surface.  In this regard,
 the Gutzwiller projected Fermi sea is known to have excellent variational energy for the $J_2-J_4$ spin model on the triangular lattice, which is believed to be appropriate for the aforementioned triangular
lattice compounds\cite{motrunich05}.

As mentioned in the introduction, the main task of this review is to illustrate the extraction of the universal properties of a QSL from the ground state wavefunction. Therefore, we now turn
to the technical challenge of calculating EE of Gutzwiller projected wavefunctions that are candidates for QSL.

\section{From QSL Wavefunctions to Entanglement Entropy} \label{sec:wfnee}

In this section, using a Monte Carlo scheme, we outline how to obtain the entanglement entropy of a ground state function that has the form of a Gutzwiller projected Slater determinant. As mentioned in the previous sections, such wavefunctions are a natural candidate for QSLs. We will focus on the second Renyi entropy $S_2$ (see Eqn. \ref{eq:defee}) since it suffices to establish topological order and is easiest to calculate within our scheme.

Let  $\phi(a,b)$ be the wavefunction of interest, where $a$ ($b$) be the configuration of subsystem A (B). Then the reduced density matrix is
$\rho_A(a,a')=\sum_b\phi^*(a,b)\phi(a',b)/N$ where the normalization $N=\sum_{a,b}\phi^*(a,b)\phi(a,b)$. The second Renyi entropy $S_2$ is:
\bea 
e^{-S_2} & =  & {\rm tr} \rho_A^2 \nonumber \\ 
  & = & \frac1{N^2}\sum_{aa'bb'} \phi^*(a,b)\phi(a',b) \phi^*(a',b')\phi(a,b')
\eea
One way to interpret this expression is to consider two copies of the system, in configuration $|a,b\rangle|a',b'\rangle$. Then, the product wavefunction
$|\Phi\rangle = \sum_{a,b,a',b'} \phi(a,b)\phi(a',b')|a,b\rangle|a',b'\rangle$ appears in the expression above. In addition if we define the ${ Swap}_A$ operator,
following Ref \cite{hastings2010}, which swaps the configurations of the $A$ subsystem in the two copies:${Swap}_A |a,b\rangle|a',b'\rangle = |a',b\rangle|a,b'\rangle$ then Renyi entropy can be written simply as:
\begin{equation}
e^{-S_2} =\frac{ \langle \Phi|{Swap}_A |\Phi\rangle}{\langle \Phi|\Phi\rangle}
\label{swap}
\end{equation}
The above expression suggests a Variational Monte Carlo algorithm for the Renyi Entropy.

Defining configurations $\alpha_1=a,\,b$, $\alpha_2=a',\,b'$ and $\beta_1=a',\,b$, $\beta_2=a,\,b'$, and $\phi_{\alpha_1}=\phi(a,b)$ etc., Eqn.(\ref{swap}) can be rewritten in the suggestive form:

\begin{equation}
\left\langle Swap_{A}\right\rangle  =  \underset{\alpha_{1}\alpha_{2}}{\sum}\rho_{\alpha_{1}}\rho_{\alpha_{2}}f\left(\alpha_{1},\alpha_{2}\right)
\end{equation}
Here the weights $\rho_{\alpha_{i}}= |\phi_{\alpha_i}|^2/{\sum_{\alpha_i}|\phi_{\alpha_i}|^2}$
are non-negative and normalized. The quantity $f\left(\alpha_{1},\alpha_{2}\right)=\frac{\phi_{\beta_1}\phi_{\beta_2}}{\phi_{\alpha_1}\phi_{\alpha_2}}$
when averaged over Markov chains generated with the probability distribution $\rho(\alpha_1)\,\rho(\alpha_2)$ yields $\left\langle Swap_{A}\right\rangle $. The Gutzwiller projected wavefunctions
to describe the QSLs we consider in this review can be written as products of determinants, as explained in the previous section. Such wavefunctions can be efficiently evaluated, which forms the basis
for Variational Monte Carlo technique \cite{gros1989}.

As shown in Ref.~\onlinecite{frank2011crit, frank2011topo}, the above algorithm, with slight modifications that we will discuss briefly in Sec.\ref{sec:gapless}, can be used to calculate $S_2$ within a few percent of error bar for subsystems with linear size $L_A \lesssim 10$,
embedded in systems with linear size $L \lesssim 20$.

\section{Gapped QSLs and Entanglement} \label{sec:gapped}

Model systems such as Toric code \cite{hamma2005} and quantum dimer models \cite{furukawa2007} provide a testing ground for the validity of Eqn.\ref{eq:teedef}. More realistic interacting quantum systems pose a greater challenge since the analytical methods are limited. Some of the earlier numerical work to extract TEE included exact diagonalization studies on small systems, looking at quantum Hall Laughlin states \cite{haque2007} and perturbed Kitaev toric code models \cite{hamma2008}. In the context of QSLs, a recent quantum Monte Carlo study \cite{isakov2011} on a sign problem free Hamiltonian used TEE to detect $Z_2$ topological order. This was a positive definite wave-function, with U(1) rather than SU(2) spin symmetry. In the rest of this section, we will focus on $SU(2)$ symmetric spin-liquids \cite{frank2011topo}, using the technique explained in the previous section.

\subsection{Establishing Topological Order in Gapped QSLs} \label{sec:establish}

In this subsection we discuss few examples where using the algorithm to calculate the entanglement entropy discussed in the previous section, one can establish topological order in 
$SU(2)$ symmetric QSLs \cite{frank2011topo}.
As a first example, let us consider an $SU(2)$ symmetric lattice wavefunction, the Kalmeyer-Laughlin Chiral Spin Liquid (CSL).

\begin{table}
\begin{tabular}{|l|l|l|}
  \hline
  State & Expected $\gamma$ & $\gamma_{\rm calculated}/\gamma_{\rm expected}$ \\
    \hline
  Unprojected ($\nu=1$) & 0 &  -0.0008$\pm$ 0.0059 $^{*}$ \\
  Chiral SL L$_A$=3  & $\log\sqrt{2}$ & 0.99 $\pm$  0.03\\
  Chiral SL L$_A$=4  & $\log\sqrt{2}$ & 0.99 $\pm$ 0.12 \\
  Lattice $\nu=1/3$   & $\log\sqrt{3}$ & 1.07$\pm$ 0.05 \\
  Lattice $\nu=1/4$    & $\log\sqrt{4}$ & 1.06 $\pm$ 0.11\\ \hline
%  Z$_2$ SL L$_A$=4 & $\log{2}$ & 0.85 $\pm$ 0.13 \\ \hline
 
\end{tabular}
\caption{Topological Entanglement Entropy $\gamma_{\textrm{calculated}}$ of spin-liquid states obtained using Monte Carlo technique discussed in the main text. $\gamma_{\textrm{expected}}$ is the exact value expected in the limit $\xi/l \rightarrow 0$ where $\xi$ is the correlation length and $l$ is the subsystem size.}
 \label{table1}
\end{table}

The Chiral SL is a spin SU(2) singlet ground state,
that breaks time reversal and parity symmetry\cite{kalmeyer,thomale}. A wavefunction in
this phase wave function may be obtained using the slave-particle
formalism by Gutzwiller projecting a $d+id$ BCS state
\cite{kalmeyer}. Alternately, it can be obtained by Gutzwiller
projection of a hopping model on the square lattice. This model has
fermions hopping on the square lattice with a $\pi$ flux through
every plaquette and imaginary hoppings across the square lattice diagonals:

\begin{eqnarray}
H=\underset{\left\langle ij\right\rangle
}{\sum}t_{ij}f_{i}^{\dagger}f_{j}+ i\underset{\left\langle
\left\langle ik\right\rangle \right\rangle
}{\sum}\Delta_{ik}f_{i}^{\dagger}f_{k} \label{Ham}
\end{eqnarray}

Here $i$ and $j$ are nearest neighbors and the hopping amplitude
$t_{ij}$ is $t$ along the $\hat y$ direction and alternating between
$t$ and $-t$ in the $\hat x$ direction from row to row; and $i$ and $k$ are second
nearest neighbors connected by hoppings along the square lattice diagonals, with amplitude
$\Delta_{ik} = i\Delta$ along the arrows and $\Delta_{ik} = -i\Delta$ against the arrows, see
Fig. 1. The unit cell contains two sublattices $A$ and $B$. This
model leads to a gapped state at half filling and the resulting
valence band has unit Chern number. This hopping model is equivalent
to a $d+id$ BCS state by an $SU\left(2\right)$ Gauge
transformation\cite{ludwig1994}. We use periodic boundary conditions
throughout this section.

\begin{figure}
\begin{centering}
\includegraphics[scale=0.35]{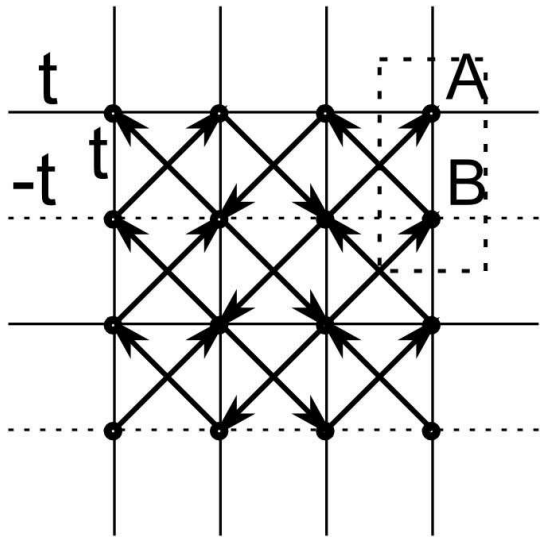}
\par\end{centering}
\caption{Illustration of a square lattice hopping model connected
with a $d+id$ superconductor. While the nearest neighbor hopping is
along the square edges with amplitude $t$ ($-t$ for hopping along
dashed lines), the second nearest neighbor hopping is along the square
diagonal (arrows in bold), with amplitude $+i\Delta$ ($-i\Delta$)
when hopping direction is along (against) the arrow. The two
sublattices in the unit cell are marked as $A$ and $B$.} \label{fig1}
\end{figure}

Due to the
fact that this Hamiltonian contains only real bipartite hoppings and
imaginary hoppings between the same sublattices and preserves the
particle-hole symmetry, this wavefunction
$\left\langle\alpha|\Phi\right\rangle$ can be written as a product
of two Slater determinants $\mathcal M Det(M_{ij})^2$, where $\mathcal M$
is just an unimportant Marshall sign factor, and:

\begin{eqnarray}
M_{ij}&  = & \left\{\left[\psi_A(k_i)+\psi_B(k_i)\right] \right. \nonumber \\    
& & \left. +(-1)^{y_j}\left[\psi_A(k_i)-\psi_B(k_i)\right]\right\}e^{ik_i \cdot r_j}
\end{eqnarray}

 \noindent  Here $\psi_A(k)$($\psi_B(k)$) is the wavefunction amplitude on sublattice A(B), $r_j$ is the coordinates of the up spins in configuration $\alpha$,
and $k_i$ is the momentums in the momentum space. The Renyi entropy
$S_{2}$ of this wavefunction can be calculated by VMC method detailed in the last
section.

\begin{figure}
\begin{centering}
\includegraphics[scale=0.4]{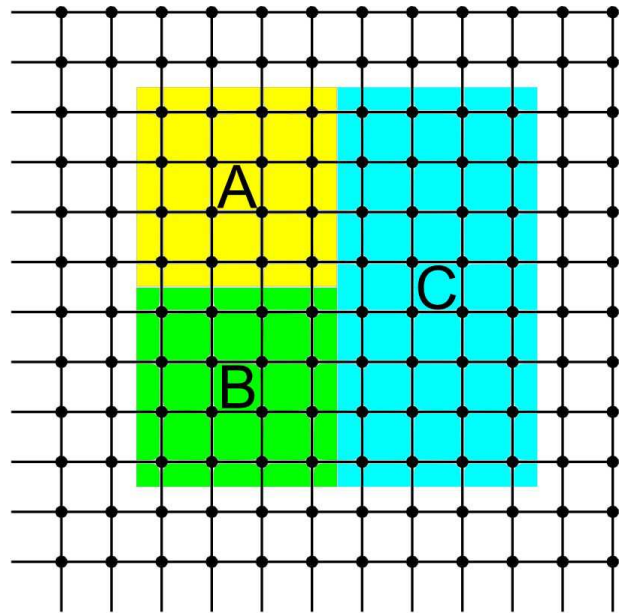}
\par\end{centering}
\caption{The separation of the system into subsystem $A$, $B$, $C$
and environment, periodic boundary condition is employed in both
$\hat{x}$ and $\hat{y}$ directions.} \label{fig3a}
\end{figure}

Since TEE $\gamma$ is a subleading term in EE, naively it is rather difficult to extract since the leading boundary law term dominates the total entropy. However, as shown
by Kitaev and Preskill \cite{kitaev2006}, one can divide the total system into four regions $A,B,C,D$ and consider the following linear combination of entropies that completely cancels out the leading non-universal term:

\begin{eqnarray}
-\gamma&=& S_{2,A}+S_{2,B}+S_{2,C}-S_{2,AB}-S_{2,AC}-S_{2,BC}+ \nonumber \\ 
& & S_{2,ABC} 
\end{eqnarray}
where $S_{2,\cal{R}}$ is the Renyi entropy $S_2$ corresponding to the region $\cal{R}$. Note that the above formula neglects terms of $O(e^{-\frac{l'}{\xi}})$ where $\xi$ is the correlation length and $l' = \textrm{min}(l_A,l_B, l_C) $. This is because the above combination of entropies cancels out all local contributions to the entanglement entropy which are expected to form a taylor series expansion in the variable $\xi/l'$, following the arguments in Ref.\cite{grov2011}. One could still have terms that are non-perturbative in  $\xi/l'$ which do not cancel out, hence the exponential dependence of the error made. This has also been confirmed by a direct calculation on select models in Ref.\cite{fradkin_raman}. Therefore, to accurately determine TEE $\gamma$, it is important that
$\xi \ll l'$.  Since we have a variational parameter $\Delta/t$ ar our disposal, one may tune it to minimze the correlation length $\xi$.
Chosing an optimal value $\Delta=0.5t$, and dividing the total system with dimensions 12 $\times$ 12 lattice spacings
into $L_A\times L_A$ squares A and B, and an $L_A\times2L_A$ rectangle C (see Fig \ref{fig3a}), one finds $\gamma=0.343\pm0.012$ for an $L_A=3$ system and $\gamma=0.344\pm0.043$ for an $L_A=4$ system \cite{frank2011topo}. Both are
in excellent consistence with the expectation of
$\gamma=\log\left(\sqrt{2}\right)=0.347$ for its ground states' two fold degeneracy.

To confirm the role of Gutzwiller projection in obtaining the QSL, one may also study the BCS wavefunction used to obtain CSL without Gutzwiller projection. This is an exactly solvable problem
and one indeed finds a value $\approx 0$ for TEE. This also serves as a benchmark for the VMC calculation. For an $L_A=3$ system,
the VMC calculation gives $\gamma=-0.0008\pm0.0059$, in agreement with vanishing TEE \cite{frank2011topo}.

\begin{figure}
\begin{centering}
\includegraphics[scale=0.4]{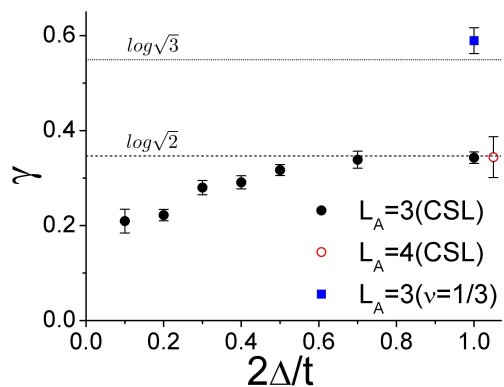}
\par\end{centering}
\caption{Illustration of finite size effect: chiral SL TEE $\gamma$ as a
function of the relative gap size given by $2\Delta/t$ for
characteristic system length $L_A=3$. The larger the relative gap, the closer the ideal value is approached. For comparison, $\gamma$ for chiral SL at $L_A=4$, and a large relative gap, is also shown. On the same plot, TEE of a lattice version of $\nu=1/3$ Laughlin state at $L_A=3$, $2\Delta/t=1.0$ is also shown. The dashed lines are the ideal TEE values of $\gamma = \log\left(\sqrt{2}\right)$ for the chiral SL and $\log\left(\sqrt{3} \right)$ for the $\nu=1/3$ Laughlin state.}
\label{fig6}
\end{figure}

One may also study the effect of increasing  correlation length on the TEE results. As mentioned above, as correlation length increases, he finite size
effects from subleading terms become more important. Indeed, as seen in Fig.\ref{fig6}, the TEE calculated using VMC deviates monotonically from the
from the universal value of $\gamma
=\log\left(\sqrt{2}\right)$ as we lower the gap size $\Delta/t$ for a system
with typical length scale $L_A=3$.

As mentioned above, the CSL ground state can be written $\sim Det(M_{ij})^2$ where $M_{ij}$ is the Slater determinant corresponding to a Chern insulator. This leads to generalization to wavefunctions that are cube or the fourth power of a Chern insulator wavefunction. For example, consider the wavefunction:

\begin{equation}\Psi_{1/3}(r_1,\,r_2\dots,\,r_N)=\Phi^3(r_1,\dots, \,r_N)\end{equation}

\noindent  where $\Phi$ is the Chern insulator Slater determinant defined above. Clearly, the product is a fermionic wavefunction, since exchanging a pair of particles leads to a change of sign.
This is similar in spirit to constructing the corresponding Laughlin
liquid of $m=3$ of fermions, by taking the cube of the Slater determinant wavefunction  in the lowest Landau level \be \psi(z_1,\dots,\,z_N)=\prod_{i<j}(z_i-z_j)e^{-\sum_i\frac{|z_i|^2}{4l^2_B}} \ee 

\noindent However, unlike the canonical Laughlin state, composed of lowest Landau level states, these are lattice wavefunctions. An interesting question is whether the lowest Landau level structure is
important in constructing states with the topological order of the Laughlin state, or whether bands with identical Chern number is sufficient, as suggested by field theoretic arguments.

Calculating the TEE using VMC with $L_A=3$ one finds $\gamma=0.5894 \pm 0.0272$ for the $m=3$ wavefunction,
consistent with the ideal value $\gamma=\log\left(\sqrt{3}\right)=0.549$. Similarly, for the fourth power of the Chern insulator Slater determinant, one finds $\gamma=0.732\pm0.076$, again consistent with ideal value in the thermodynamic limit:
  $\gamma=\log\left(\sqrt{4}\right)=0.693$.

These results offer direct support for the TEE formula
$\gamma=\log D$ as well as their validity as topological ground
state wavefunctions carrying fractional charge and statistics. The
lattice fractional Quantum Hall wavefunctions discussed here may be
relevant to the recent studies of flat band Hamiltonians with
fractional quantum Hall states \cite{tang2010, neupert2010, sun2010,
sheng2011}.

\subsection{Beyond TEE: Extracting Quasiparticle Statistics Using Entanglement} \label{sec:topodata}

At the fundamental level, topologically ordered phases in two dimensions display a number of unique properties such as nontrivial statistics
of emergent excitations. For example, in Abelian topological phases, exchange of identical excitations or taking one excitation around another (braiding) leads
 to characteristic phase factors, which implies that these excitations are neither bosonic nor fermionic.  A further remarkable generalization of statistics occurs in non-Abelian phases where excitations introduce
 a degeneracy. An important and interesting question is whether the ground state directly encodes this
 information, and if so how one may access it? It is well known that topologically ordered phases feature a ground state degeneracy that depends on the topology of the space on
 which they are defined. Also, as discussed in the previous sections, the ground state wavefunctions of such states contain a topological contribution to the quantum entanglement, the topological entanglement entropy (TEE).
In this section we show that
combined together, these two ground state properties can be used to extract the generalized statistics associated with excitations in these states.

The generalized statistics of quasiparticles is formally captured by the modular $\mathcal S$ and $\mathcal{U}$
matrices, in both Abelian and non-Abelian states  \cite{wen1990,wen93,kitaev_honey,nayak,yellowbook}.
The element $\mathcal{S}_{ij}$ of the modular $\mathcal S$ matrix determines the mutual statistics of $i$'th
quasiparticle with respect to the $j$'th quasiparticle while the element $\mathcal{U}_{ii}$ of (diagonal) $\mathcal{U}$ matrix determines the self-statistics (`topological spin') of the $i$'th quasiparticle.
Note, these provide a nearly complete description of a topologically ordered phase - for instance, fusion rules that dictate the outcome of bringing together a pair of quasiparticles, are determined from the
modular $\mathcal S$ matrix, by the Verlinde formula\cite{Verlinde}. Previously, Wen proposed \cite{wen1990} using the non-Abelian Berry phase to extract statistics of quasiparticles. However, the idea in
Ref.~\onlinecite{wen1990} requires one to have access to an \textit{infinite set of ground-states}
labeled by a continuous parameter, and is difficult to implement. Recently, Bais et al. \cite{Bais} also discussed extracting $\mathcal S$ matrix in numerical simulations, by explicit braiding of
 excitations. In contrast, here we will just use the set of ground states on a torus, to determine the braiding and fusing of gapped excitations.

\begin{figure}
\begin{centering}
\includegraphics[scale=0.35]{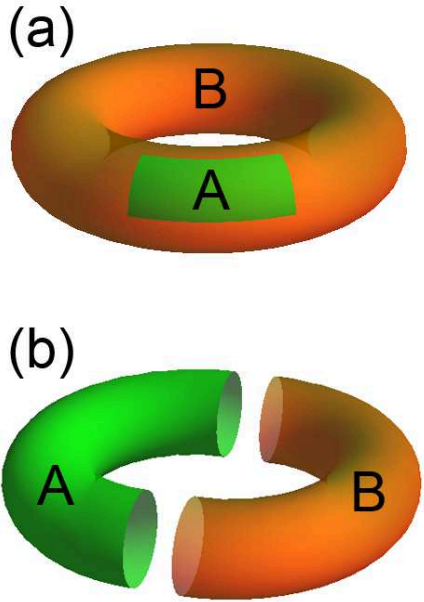}
\par\end{centering}
\caption{Two types of entanglement bipartitions on the torus: (a) A
trivial bipartition with contractible boundaries for which the TEE
$\gamma = \log D$, and (b) A bipartition with non-contractible
boundaries, where the TEE depends on ground state. } \label{fig8}
\end{figure}

It is sometimes stated without qualification, that TEE is a quantity solely determined by the total quantum dimension $D$ of the underlying topological theory.
However, this holds true \textit{only when the
boundary of the region $A$ consists of topologically trivial closed
loops}. If the boundary of region $A$ is non-contractible, for
example if one divides the torus into a pair of cylinders, \textit{generically the
entanglement entropy is different for different ground states} (see Figure \ref{fig8}). Indeed
as shown in Ref.~\onlinecite{dong2008} for a class of topological
states, the TEE depends on the particular linear combination
of the ground states when the boundary of region $A$ contains
non-contractible loops. We will exploit this dependence to extract information about the topological phase beyond the total quantum dimension $D$.

Let us briefly summarize the key ideas involved in extracting braiding statistics using quantum entanglement \cite{frank2012} and then discuss the details.
We recall that the number of ground states $N$ on a  torus corresponds to the number of distinct quasiparticle types \cite{wen1990, wen1991, footnote_gsd}. Intuitively,
 different ground states are generated by inserting appropriate fluxes `inside' the cycle of the torus, which is only detected by loops circling the torus. We would like to express these
 quasiparticle states as a linear combination of ground states. A critical insight is that this can be done using the topological entanglement entropy for a region $A$ that wraps around
 the relevant cycle of the torus. To achieve this, we introduce the notion of \emph{minimum entropy
states (MESs)} $\Xi_j$, $j=1-N$, namely the ground states with minimal entanglement
entropy (or maximal TEE, since the TEE always
reduces the entropy) for a given bipartition. These states are
generated by insertion of a definite quasi-particle into the cycle enclosed by region A. With this in hand, one can readily access the modular $\mathcal S $ and $\mathcal U$ matrices as we now describe.

\subsubsection{Extracting Statistics from Topological Entanglement Entropy} \label{sec:stat}

As mentioned above, for Abelian phases, the
$ij$'th entry of the $\mathcal{S}$ matrix corresponds to the phase
the $i$'th quasi-particle acquires when it encircles the $j$'th
quasi-particle. The $\mathcal{U}$ matrix is diagonal and the $ii$'th
entry corresponds to the phase the $i$'th quasi-particle acquires
when it is exchanged with an identical one. Since the MESs are the
eigenstates of the nonlocal operators defined on the entanglement
cut, the MESs are the canonical basis for defining $\mathcal S$ and
$\mathcal U$. The modular matrices are just certain unitary
transformations of the MES basis. For example,
\begin{equation}
\mathcal{S}_{\alpha\beta}=\frac{1}{D}\left\langle
\Xi_{\alpha}^{\hat{x}}|\Xi_{\beta}^{\hat{y}}\right\rangle
\label{eq:defineS}
\end{equation}

Here $D$ is the total quantum dimension and $\hat{x}$ and $\hat{y}$
are two directions on a torus. Eqn.\ref{eq:defineS} is just a
unitary transformation between the particle states along different
directions. In the case of a system with square geometry, the $\mathcal S$ matrix acts as a $\pi/2$ rotation on
the MES basis $\left|\Xi_{\beta}^{\hat{y}}\right\rangle$. In
general, however, $\hat{x}$ and $\hat{y}$ do not need to be
geometrically orthogonal, and the system does not need to be
rotationally symmetric, as long as the loops defining
$\left|\Xi_{\alpha}^{\hat{x}}\right\rangle$ and
$\left|\Xi_{\beta}^{\hat{y}}\right\rangle$ interwind with each
other. We should mention one caveat with Eqn.\ref{eq:defineS}. When we write Eqn.\ref{eq:defineS}, we assume that the quasiparticles transform trivially under rotations. However, there exist topologically ordered states where rotations can induce a change in the quasiparticle type. For example, in the Wen-Plaquette model \cite{wenplaq}, the rotations convert a charge-type excitation to a flux-type excitation. In such topological states, the procedure outlined above can give incorrect results, and we exclude such states from our discussion.

More concretely, let us denote the set of ground state wavefunctions as $|\xi_\alpha\rangle$ where
$\alpha = 1-N$ and $N$ is the total ground state degeneracy. To obtain the modular $S$ matrix, one calculates the minimum entropy states $|\Xi_{1,\alpha}\rangle$
and $|\Xi_{2,\alpha}\rangle$ for entanglement cuts along the directions $\vec{w}_1$ and $\vec{w}_2$ respectively.
Doing so leads one to the matrices $U_1$  and $U_2$ that relates the states $|\xi_\alpha\rangle$ to  $|\Xi_{1,\alpha}\rangle$  and $|\Xi_{2,\alpha}\rangle$
respectively. The modular $\mathcal{S}$ matrix is given by $U_2^{-1} U_1$ upto
an undetermined phase for each MES. The existence of an identity particle that obtains
trivial phase encircling any quasi-particle helps to fix the
relative phase between different MESs, requiring the entries of the
first row and column to be real and positive. This completely
defines the modular $\mathcal{S}$ matrix.
Therefore, the modular $\mathcal{S}$ matrix can be derived
even without any presumed symmetry of the given wave functions.

\begin{figure}
\begin{centering}
\includegraphics[scale=0.7]{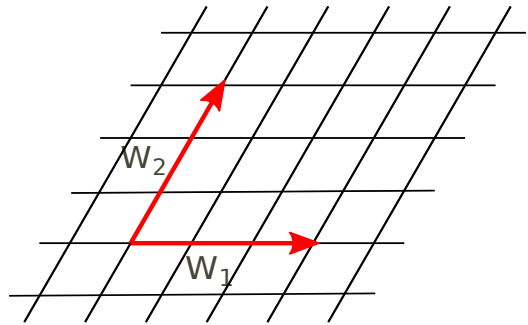}
\par\end{centering}
\caption{Vectors $\vec{w}_1$ and $\vec{w}_2$ that define a lattice with periodic boundary conditions. That is, any point $\vec{r}$ on the lattice satisfies $\vec{r} \equiv \vec{r}+\vec{w}_1 \equiv \vec{r}+\vec{w}_2$.
The area of the lattice is $|\vec{w}_1 \times \vec{w}_2|$ where $\times$ denotes the cross product.}
\label{w1w2}
\end{figure}

The above algorithm is able to extract the modular
transformation matrix $\mathcal{S}$ and hence braiding and mutual statistics of quasi-particle
excitations just using the ground-state wave
functions as an input. Further, there is no loss of generality for
non-Abelian phases, which can be dealt by enforcing the orthogonality condition in step
$2$ which guarantees that one obtains states with quantum dimensions $d_\alpha$ in an increasing
order.
As shown in Ref.~\onlinecite{frank2012}, if the lattice has $2\pi/3$ rotation symmetry, then the operation of rotating MES states by an angle of
$2\pi/3$ corresponds to the action of $\mathcal{US}$ on MES. Therfore, in the presence of such a symmetry, one can extract the $\mathcal{U}$ matrix as well.
\subsubsection{Example: Semionic Statistics in Chiral Spin-Liquid from Entanglement Entropy} \label{subsec:semion}

Let us revisit Chiral Spin-Liquid (CSL), first discussed in the Sec.\ref{sec:establish}. Even though topological properties of CSL are well established using field-theoretic methods \cite{wen1990},
from a wavefunction point of view, CSL cannot be dealt with analytically unlike continuum Laughlin states. For example, the low energy theory of CSL predicts a particle with a semionic statistics (that is,
the particle picks up a factor of -1 when it goes $2\pi$ around an identical one) which is difficult to verify in
the lattice wavefunction directly. We will show that the entanglement based algorithm discussed in the last subsection readily
demonstrates the existence of semion as an excitation in CSL.

As discussed in Sec.\ref{sec:establish}, CSL corresponds to a $\textrm{d+id}$-wave state for chargeless spinons. The corresponding ground state wavefunction is obtained
by Gutzwiller projecting the mean-field BCS ground state wavefunction. CSL has a topological ground-state degeneracy of two on torus. One of the ground states can be obtained
by enforcing periodic boundary condition along $\hat{x}$-direction and anti-periodic along $\hat{y}$. We denote it by $|0,\pi\rangle$. The orthogonal ground state is given by
$|\pi,0\rangle$ with the same notation.

Our task is to obtain the modular $\mathcal{S}$-matrix since it encodes the mutual statistics of quasiparticles. To do so, as outlined in the previous subsection, we consider linear combinations
of $\left|0,\pi\right\rangle $ and
$\left| \pi,0\right\rangle$:

\begin{equation}
\left|\Phi(\phi)\right\rangle =\cos\phi\left|0,\pi\right\rangle
+\sin\phi\left|\pi,0\right\rangle
\end{equation}

\noindent  and numerically minimize the entanglement entropy with respect to the parameter $\phi$ so as to obtain the MESs \cite{frank2012}.
For system with square geometry the $\mathcal{S}$ matrix describes
the action of $\pi/2$ rotation on the MESs. Since the two states
$|0,\pi\rangle$ and $|\pi,0\rangle$ transform into each other under
$\pi/2$ rotation, this implies that $\mathcal{S}$ matrix is given by $V^{\dagger}  \sigma_x V$ where $V$ is the unitary matrix that rotates the
$\{|0,\pi\rangle\,|\pi,0\rangle\}$ basis to the MESs basis. From Fig.\ref{fig4}, one finds that

\begin{eqnarray*} V \approx \left(\begin{array}{cc}
\cos(0.14 \pi) & -\sin(0.14 \pi )e^{i\varphi} \\
\sin(0.14\pi) & \cos(0.14\pi )e^{i\varphi} \end{array} \right)
\end{eqnarray*}

The existence of an identity particle requires positive real entries
in the first row and column and implies $\varphi=0$, which gives:

\begin{eqnarray*} \mathcal{S} \approx \left(\begin{array}{cc}
0.77 & 0.63 \\
0.63 & -0.77 \end{array}\right) =  \frac{1}{\sqrt{2}}\left(\begin{array}{cc}
1.08 & 0.89 \\
0.89 & -1.08 \end{array}\right)
\end{eqnarray*}

\noindent  which is in good agreement with the exact result

\be
\mathcal{S}_{\textrm{exact}} = \frac{1}{\sqrt{2}}\left(\begin{array}{cc}
1 & 1 \\
1 & -1 \end{array}\right)
\ee
Even though the  $\mathcal{S}$ matrix
obtained using our method is approximate, the quasiparticle
statistics can be extracted exactly. In particular, the above $\mathcal{S}$
matrix tells us that the quasi-particle corresponding to $d_0 = 1$
does not acquire any phase when it goes around any other particle
and corresponds to an identity particle as expected, while the
quasi-particle corresponding to $d_{1/2} = 1$ has semion statistics
since it acquires a phase of $\pi$ when it encircles another
identical particle. Numerical improvements can further reduce the
error in pinpointing the MES and thereby leading to a more accurate
value of the $\mathcal S$ matrix.

It is also interesting to compare the full dependence of TEE on the angle $\phi$. Theoretically, one finds the
following expression for TEE \cite{dong2008}:

\begin{eqnarray}
2\gamma-\gamma' &=&\log\frac{4}{3+\sin\left(4\phi\right)}
\label{eqntee7}
\end{eqnarray}

\noindent  where $\gamma$ is TEE for a region with contractible boundary while $\gamma'$ is that for a region with non-contractible boundary. From Fig.\ref{fig4}, we see that the numerically evaluated TEE is in rather good agreement with the exact expression.

\begin{figure}
\begin{centering}
\includegraphics[scale=0.2]{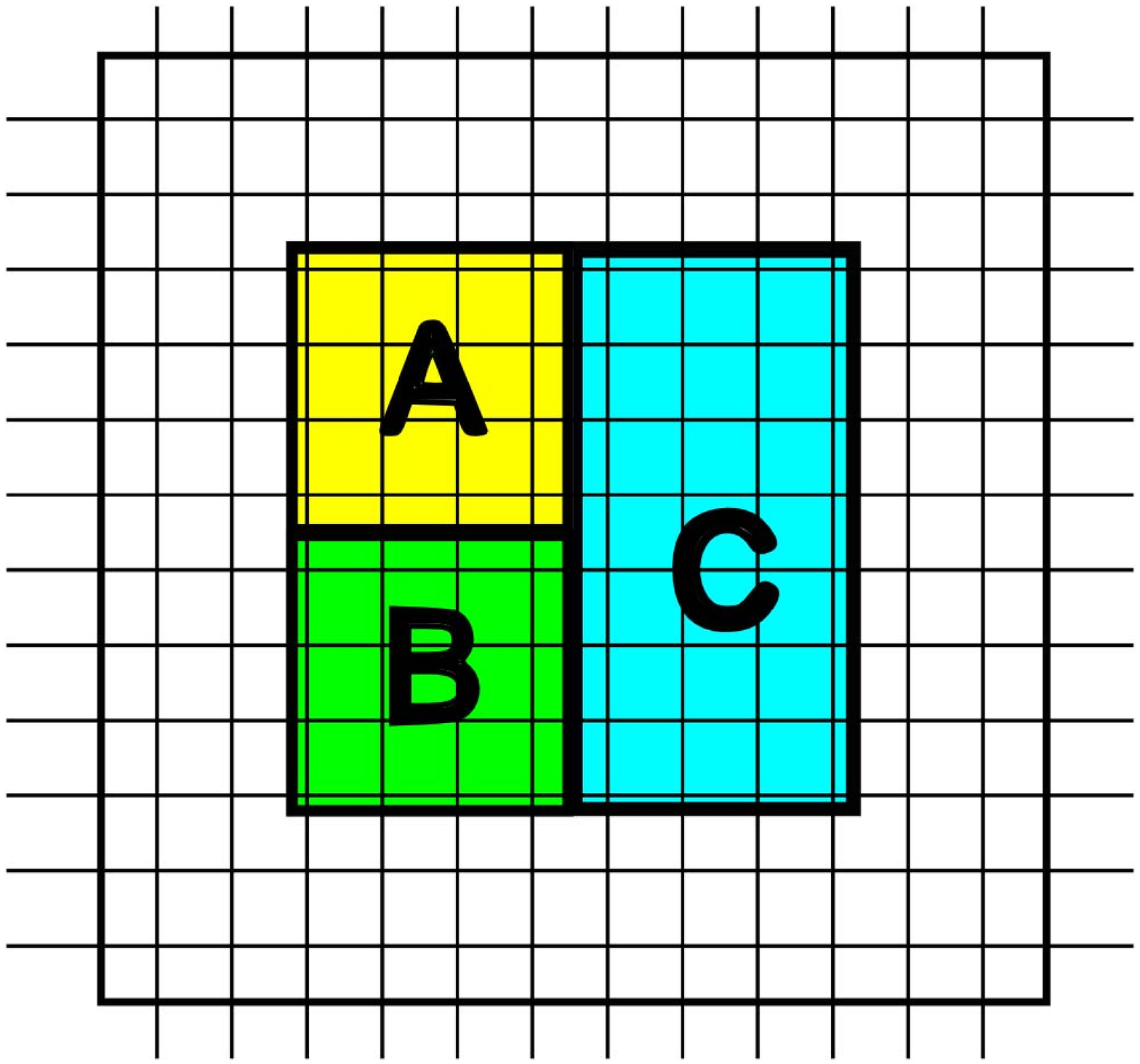}
\includegraphics[scale=0.2]{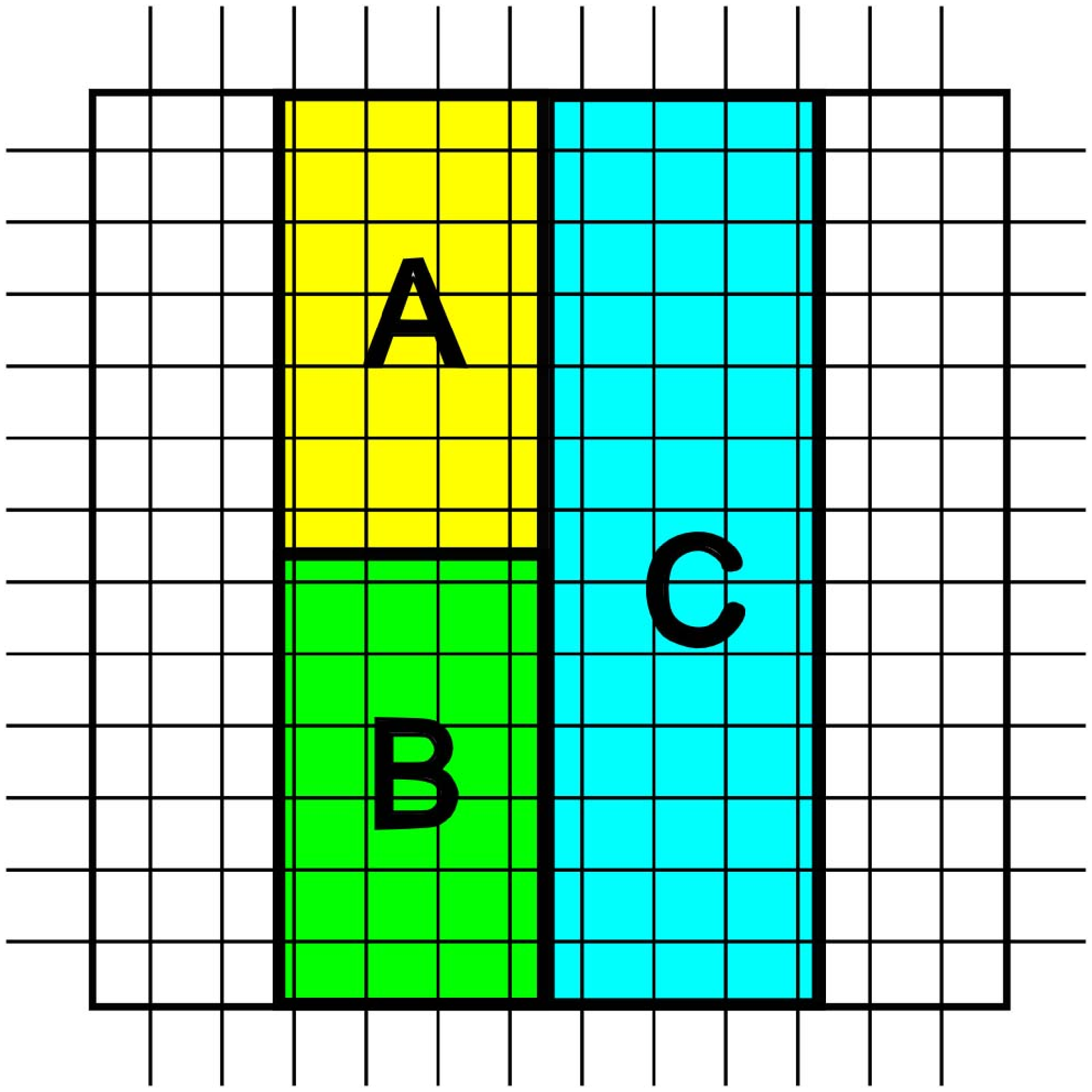}
\par\end{centering}
\caption{The separation of the system into subsystem $A$, $B$, $C$
and environment, periodic or antiperiodic boundary condition is
employed in both $\hat{x}$ and $\hat{y}$ directions. a: The
subsystem $ABC$ is an isolated square and the measured TEE has no
ground state dependence. b: The subsystem $ABC$ takes a non-trivial
cylindrical geometry and wraps around the $\hat{y}$ direction, and
TEE may possess ground stated dependence.} \label{fig3}
\end{figure}

\begin{figure}
\begin{centering}
\includegraphics[scale=0.42]{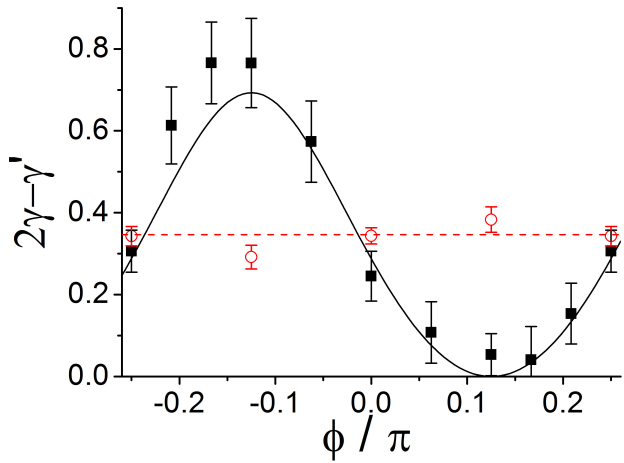}
\par\end{centering}
\caption{The black dots show the numerically measured TEE $2\gamma - \gamma'$ for a CSL
ground state from linear combination $\left|\Phi\right\rangle
=\cos\phi\left|0,\pi\right\rangle +\sin\phi\left|\pi,0\right\rangle
$ as a function of $\phi$ with VMC simulations \cite{frank2012} using geometry in
Fig. \ref{fig3}b . Here $\gamma$ is TEE for a region with contractible boundary while $\gamma'$ is that for a region with non-contractible boundary. The solid curve is the theoretical value from Eqn.
\ref{eqntee7}. The periodicity is $\pi/2$. The red dots show the TEE for the same linear combination
for a trivial bipartition. In the latter case, TEE is essentially independent of $\phi$ and again agrees rather well with the theoretical
expectation (the dashed red curve).} \label{fig4}
\end{figure}

\section{Gapless QSLs and Entanglement} \label{sec:gapless}

In this section we consider Gutzwiller projected Fermi liquid wave functions which are considered good ansatz for
ground states of critical spin-liquids \cite{motrunich05}. We analyze two different classes of critical spin-liquids: states that at the slave-particle mean-field level have a
full Fermi surface of spinons and those with only nodal spinons. The basic idea we use here is that for free fermions, the boundary law is violated by a multiplicative
logarithm, that is, $S(l) \sim l^{d-1} \log(l)$ \cite{gioev2006, wolf}. This suggests that for critical spin-liquids, whose low-energy theory contains a spinon Fermi surface \cite{mattergauge}, a similar result might be expected.
Indeed, one finds that for such Gutzwiller projected wavefunctions the entanglement entropy scales as $S(l) \sim l \log(l)$ for small system sizes \cite{frank2011crit}.
On the other hand, for wavefunctions with nodal spinons, the low-energy theory is believed to a two-dimensional CFT for which the result in Eqn.\ref{eq:twodcft} must hold. Thus, in this case, the VMC algorithm
can be used to extract the universal shape dependent subleading constant.

%Consider a mean-field state of spinons hopping on  a triangular lattice with uniform hopping $t_{rr'} = t$. Thus, one obtains a Fermi surface of spinons at the mean-field level while for a
%  square-lattice with $\pi$
%flux through every plaquette (i.e. $\Pi_{\Box} t_{rr'} = -1$) one obtains nodal Dirac spinons. We also study the projected wave function on square lattice with uniform hopping (and no flux).

The wave-functions for these spin-liquids are constructed by starting with a system of spin-$1/2$
fermionic spinons $f_{r\alpha}$ hopping on a finite lattice at half-filling with a mean field Hamiltonian:$
 H_{MF} = \sum_{r r'} \left[  -t_{r r'}f^{\dagger}_{r\sigma} f_{r'\sigma} + h.c. \right]$. The spin wave-function is given by $|\phi\rangle = P_G |\phi\rangle_{MF}$ where $|\phi\rangle_{MF}$ is the ground state
 of $H_{MF}$ and the Gutzwiller projector
$P_G = \prod_i \left( 1- n_{i \uparrow} n_{i \downarrow}\right) $ ensures exactly one fermion per site. The one dimensional case was discussed in Ref.~\onlinecite{Cirac}.

%The sign-structure of the projected wave-function depends markedly on the underlying lattice.
%For a bipartite lattice with $t_{r r'}$ non-zero and real only for the opposite sublattices, one can prove that the wave-function satisfies the Marshall sign rule.
%
%Therefore, for a bipartite lattice, one only needs to calculate
%$\langle Swap_{A,mod}\rangle$ since $\langle Swap_{A,sign}\rangle$ trivially equals unity. Thus the projected wave-function for the square lattice with and without $\pi$-flux (as well as that for the
%one-dimensional Haldane-Shastry state) satisfies the Marshall's sign rule while that for the triangular lattice doesn't.
%We discuss these three cases separately. 

\subsection{Critical QSL with Spinon Fermi Surface}

\begin{figure}[tb]
 \centerline{
   \includegraphics[width=240pt, height=240pt]{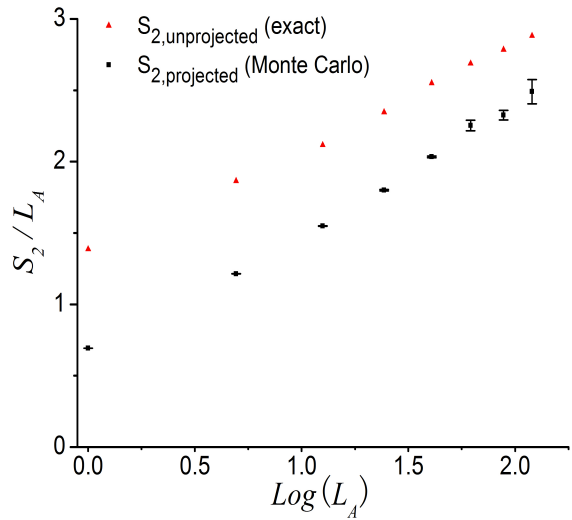}}
 \caption{Renyi entropy data for the projected and unprojected Fermi sea state on the triangular lattice of size $18 \times 18$ with $L_A = 1-8$. Note, projection barely modifies the slope,
 pointing to a Fermi surface in the spin wavefunction.}
 \label{tri}
 \end{figure}

Consider the mean-field ansatz that describes spinons hopping on a triangular lattice. In this case, one might expect that the projected wavefunction describes the ground state of a QSL whose low energy theory contains a spinon Fermi surface \cite{motrunich05, mattergauge}. Indeed, one finds \cite{frank2011crit} that for a triangular lattice of total size $18 \times 18$
on a torus and a subregion $A$ of square geometry with linear size $L_A$ up to 8 sites, the Renyi entropy scales as  $L_A \,\textrm{log} L_A$  (Fig. \ref{tri}).
This is rather striking since the wave-function is a spin wave-function and therefore could also be written in terms of hard-core bosons. This result strongly suggests the presence of an underlying
 spinon Fermi surface. In fact the coefficient of the $L_A \,\textrm{log} L_A$ term is rather similar before and after projection. This observation may be rationalized as follows. Consider one dimensional fermions with $N$ flavors, where the central charge (and hence the coefficient of the $\log L_A$ entropy term) are proportional to $N$. However, Gutzwiller projection
 to single occupancy reduces the central charge to $N-1$. This is a negligible change for large $N$. Similarly, a two dimensional Fermi surface may be considered a collection of many
  independent one dimensional systems in momentum space, and Gutzwiller projection removes only one degree of freedom.

 \begin{figure}[tb]
 \centerline{
   \includegraphics[width=240pt, height=230pt]{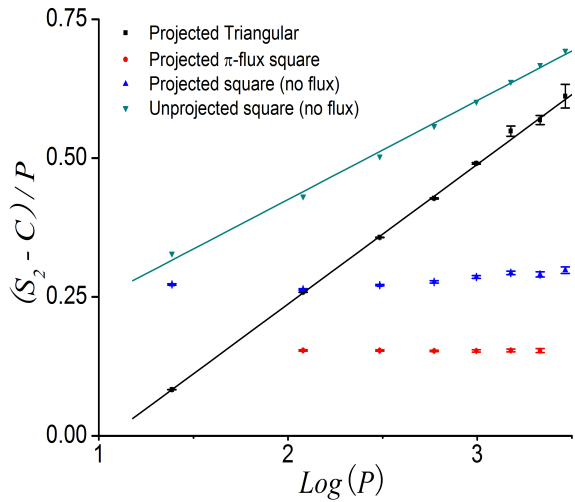}}
\caption{Comparison of the Renyi entropy data for the projected Fermi sea state on the triangular lattice and square (with and without $\pi$-flux) lattice as a function of the perimeter $P$
of the subsystem $A$. Here $C$ is the constant part of the $S_2$. We find $S_2 \sim P log(P) + C$ for the projected triangular lattice state while $S_2 \sim P + C$ for the projected $\pi$-flux square
 lattice state. For the square lattice state (no flux), the projection leads to a significant reduction in $S_2$ and the data suggests at most a very weak violation of the area-law in the projected state.}
 \label{all}
 \end{figure}
The area-law violation of the Renyi entropy
for Gutzwiller projected wave-functions substantiates the theoretical expectation that an underlying Fermi surface is present in the spin wavefunction.

\subsection{Critical QSL with Nodal Spinons}

Next, consider a mean-field ansatz where the spinons hop on a square lattice with a flux $\pi$ through every plaquette \cite{piflux}. Thus, at the mean-field level, one obtains spinons with Dirac dispersion around two nodes, say, $(\pi/2,\pi/2)$ and $(\pi/2,-\pi/2)$ (the location of the nodes depend on the
gauge one uses to enforce the $\pi$ flux). The projected wave-function has been proposed in the past as the ground state of an algebraic spin liquid. The algebraic spin-liquid
is believed to be describable by a strongly coupled conformal field theory of Dirac spinons coupled to a non-compact $SU(2)$ gauge field \cite{wen2002, wen2004}. Because of this
the algebraic spin-liquid has algebraically decaying spin-spin correlations which can be verified explicitly using Variational Monte Carlo \cite{frank2011crit}.

Square lattice being bipartite, the projected wavefunction satisfies Marshall's sign rule and hence one is able to perform Monte Carlo calculation of Renyi entropy on bigger lattice sizes
compared to the triangular lattice case. Consider an overall geometry of a torus of size $L_A \times 4 L_A$, $L_A \leq 14$, with both region $A$ and its complement of sizes  $L_A \times 2 L_A$
(the total boundary size being $L_A + L_A = 2 L_A$). One finds \cite{frank2011crit} that the projected wavefunction follows an area law akin to its unprojected counterpart
and has the scaling $S_2 \approx 0.31 L_A  + \gamma$ where $\gamma \approx 1.13$ is a universal constant that depends only on the aspect ratio of the geometry \cite{ryu2006}.
% This is consistent with the presence of Dirac fermions at low energies.

\subsection{Effect of Projection on Nested Fermi Surface}
Finally, consider a mean-field ansatz as the fermions on a square lattice with nearest-neighbor hopping. The unprojected Fermi surface is nested. Since projection amounts to taking correlations into account, one might wonder whether the Fermi surface undergoes a magnetic instability after the projection. Indeed,
 one finds that the non-zero magnetic order in the projected wave-function \cite{frank2011crit}, consistent with an independent  study Ref.~\onlinecite{li2011}. This was verified by calculating spin-spin correlations on a
 $42 \times 42$ lattice. Renyi entropy results for a a lattice of total size $24 \times 24$ with region $A$ being a square up to size $12 \times 12$ are shown in the Fig. \ref{all}.
Though it is difficult to rule out presence of a partial Fermi surface from Renyi entropy, there is a significant reduction in the Renyi entropy as well as the coefficient of $L_A log L_A$ term
as compared to the unprojected Fermi sea.

\section{Synopsis and Recent Developments} \label{sec:synop}

In this article, we reviewed the perspective provided by the concept
of entanglement entropy on QSLs. In particular, we discussed the
practical uses of the concept of TEE to detect topological order in
realistic QSLs and a detailed characterization of topological order
using entanglement as well. Returning to the questions posed in the
introduction, let us discuss whether entanglement entropy is of any
value in finding new QSLs? Remarkably, one of the most accurate
methods to study strongly correlated systems, namely, Density Matrix
Renormalization Group (DMRG) \cite{white1992}, employs entanglement entropy as a
resource to find ground states of particular Hamiltonians. In brief,
DMRG implements a real-space renormalization group that successively
coarse-grains the system on the basis of the entanglement between a
sub-region and the rest of the system. Though DMRG is most accurate
for one-dimensional quantum systems, recent works are now beginning
to explore the realm of two-dimensional strongly correlated systems.
 As an output, DMRG yields an approximate ground state of the system along with the entanglement entropy for a
bipartition of the system into two halves. Having access to the entanglement entropy implies that one can use the results reviewed in the previous sections, thus allowing one to establish (or rule out)
a QSL ground state. As an example, using DMRG Yan et al \cite{yan2010} found that the ground state of 
kagome lattice Heisenberg Hamiltonian is a QSL. In a more recent study, Jiang et al \cite{hongchen2012} established the topological nature of this spin-liquid by calculating the TEE of 
the ground state for a region $A$ with non-contractible boundary and found it to be close to $\log(2)$, a value that is apparently consistent with that of a $Z_2$ QSL. Similar results are found for other models which are known to have topologically ordered ground states \cite{hongchen2012}. However, as discussed in Sec.\ref{sec:stat}, TEE for a region with non-contractible boundary can be any number between 0 and $\log(2)$, the maximum value of TEE (hence minimum value for total entropy),being attained when the ground state corresponds to quasiparticle states. These results indicate that DMRG might be biased towards a linear combination of ground states that minimize total entropy.  
As discussed  in Sec.\ref{sec:topodata},  entanglement can be used to determine quasiparticle statistics from a topologically degenerate set of ground states. Recently, using such an approach, Cincio and Vidal \cite{cincio}  extracted the modular $S$ and $U$ matrices for a microscopic lattice Hamiltonian 
that has a Chiral Spin Liquid ground state. Entanglement based measures have therefore emerged as a powerful conceptual and numerical tool in the context of quantum spin liquids. One may expect that in future, the confluence of ideas of many body physics and quantum information will throw up other equally fruitful directions.

\textbf{Acknowledgements:} We thank Ari Turner and Masaki Oshikawa for useful discussions and collaboration on the topics in this review. Support from NSF DMR- 0645691 and hospitality at KITP, Santa Barbara where part of this work was performed, is gratefully acknowledged. This research was supported in part by the National Science Foundation under Grant No. NSF PHY11-25915.


\begin{thebibliography}{7}
\bibitem{balentsreview} L. Balents, Nature 464, 199 (2010).
\bibitem{anderson1973} P. W. Anderson, Mater. Res. Bull. 8, 2 (1973).
\bibitem{anderson1987} P.W. Anderson, Science 237, 1196 (1987).
\bibitem{baskaran1987} G. Baskaran, Z. Zou, P. Anderson, Solid State Comm. 63 973.
\bibitem{read1991} N. Read and S. Sachdev, Phys. Rev. Lett. 66, 1773 (1991); S. Sachdev, Physical Review B 45, 12377 (1992).
\bibitem{kivelson1987} S. Kivelson, D. Rokhsar, J. Sethna, Phys. Rev. B 35, 8865-8868 (1987).

\bibitem{read1989} N. Read, B. Chakraborty, Phys. Rev. B 40, 7133 (1989).

\bibitem{senthil2004} T. Senthil, A. Vishwanath, L. Balents, S. Sachdev and M. P. A. Fisher, Science 303, 1490 (2004).
\bibitem{exp} Y. Shimizu et al. Phys. Rev. Lett. 91, 107001 (2003),
 Y. Okamoto et al. Phys. Rev. Lett. 99. 137207 (2007), M. Yamashita et al, Science 328, 1246 (2010), J. S. Helton et al., Phys. Rev. Lett. 98, 107204 (2007).
\bibitem{yan2010} Simeng Yan, David A. Huse, Steven R. White, Science 332, 1173 (2011).
\bibitem{hongchen2012} Hong-Chen Jiang, Zhenghan Wang, Leon Balents, arXiv:1205.4289.
\bibitem{Balents} H.-C. Jiang, H. Yao, L. Balents, arXiv:1112.2241.
\bibitem{Wen} L. Wang, Z.-C. Gu, F. Verstraete, X.-G. Wen, arXiv:1112.3331.
\bibitem{meng2010} Z. Y. Meng, T. C. Lang, S. Wessel, F. F. Assaad, A. Muramatsu, Nature 464, 847 (2010).
\bibitem{wen2004} Xiao-Gang Wen, \textit{Quantum field theory of many-body systems}, Oxford Graduate Texts, 2004.
\bibitem{hastings2004} M. B. Hastings, Phys. Rev. B 69, 104431 (2004).
\bibitem{mattergauge} see e.g. B. I. Halperin, P. A. Lee, and N. Read, Phys. Rev. B 47, 7312 (1993). B. L. Altshuler, L. B. Ioffe, and A. J. Millis, Phys. Rev. B 50, 14048 (1994). David F. Mross et al, Phys. Rev. B 82, 045121 (2010). Sung-Sik Lee, Phys. Rev. B 80, 165102 (2009).
\bibitem{kitaev2006} A. Kitaev, J. Preskill,  Phys. Rev. Lett. 96, 110404 (2010).
\bibitem{levin2006} M. Levin, X.-G. Wen, Phys. Rev. Lett. 96, 110405 (2006).
\bibitem{grov2011} T. Grover, A. Turner and A. Vishwanath, Phys. Rev. B 84, 075128 (2011).
\bibitem{senthilswingle} B. Swingle, T. Senthil, 	arXiv:1109.3185v1.
\bibitem{pufu} Igor R. Klebanov, Silviu S. Pufu, Subir Sachdev, Benjamin R. Safdi, JHEP 1205, 036 (2012).
\bibitem{gioev2006} D. Gioev, I. Klich, Phys. Rev. Lett. 96, 100503 (2006).
\bibitem{wolf} M. M. Wolf, Phys. Rev. Lett. 96, 010404 (2006).
\bibitem{bombelli} L. R. Bombelli. K. Koul, J. Lee, and R. D. Sorkin,  Phys.
Rev. D 34, 373 (1986).
\bibitem{srednicki} M. Srednicki, Phys. Rev. Lett. 71, 66 (1993).
\bibitem{hastingsproof} M. B. Hastings, J. Stat. Mech. , P08024 (2007).
\bibitem{eisert2010}  J. Eisert, M. Cramer, M. B. Plenio, Rev. Mod. Phys. 82, 277 (2010).
\bibitem{ryu2006} S. Ryu, T. Takayanagi, JHEP 0608:045, 2006. H.Casini, M.Huerta, Nucl. Phys. B 764, 183 (2007).
\bibitem{cftwork} P. Calabrese, J. Cardy, J. Stat. Mech. , P06002 (2004). C. Callan, F. Wilczek, Phys. Lett. B 333, 55 (1994).  Holzhey, C., F. Larsen, and F. Wilczek, Nucl. Phys. B 424, 443
(1994). 
\bibitem{hamma2005} A. Hamma, R. Ionicioiu, and P. Zanardi, Phys. Lett. A 337, 22 (2005); Phys. Rev. A 71, 022315 (2005).
\bibitem{kitaev2003} A. Kitaev, Ann. Phys. 303, 2 (2003).

\bibitem{levin2005} Michael A. Levin and Xiao-Gang Wen, Phys RevB. 71.045110 (2005).
\bibitem{HaldaneLi} Hui Li and F. D. M. Haldane , Phys. Rev. Lett. 101, 010504 (2008).

\bibitem{rokhsar1988} D. S. Rokhsar and S. A. Kivelson, Phys. Rev. Lett. 61, 2376 (1988).
\bibitem{moessner2001} R. Moessner, S. L. Sondhi, Phys. Rev. Lett. 86, 1881 (2001).
\bibitem{senthil2002} T. Senthil, O. Motrunich, Phys. Rev. B 66, 205104 (2002).
\bibitem{balents2002} L. Balents, M.P.A. Fisher, S.M. Girvin, Phys. Rev. B 65, 224412 (2002).
\bibitem{misguich} G. Misguich, Phys.Rev.Lett. 89, 137202  (2002).
\bibitem{gutzwiller1963} M. Gutzwiller, Phys. Rev. Lett. 10, 159 (1963);Phys. Rev. 134, A923 (1964);ibid. 137, A1726 (1965).
\bibitem{motrunich05} O. I. Motrunich, Phys. Rev. B 72, 045105 (2005). 

\bibitem{varstudy} Ying Ran, Michael Hermele, Patrick A. Lee, Xiao-Gang Wen, Phys. Rev. Lett. 98, 117205 (2007); Tocchio et al, Phys. Rev. B 80, 064419 (2010); T. Grover et al, Phys. Rev. B 81, 245121 (2010);  B. K. Clark, D. A. Abanin, S. L. Sondhi, Phys. Rev. Lett. 107, 087204 (2011).
\bibitem{kalmeyer} V. Kalmeyer and R. B. Laughlin, Phys. Rev. Lett. 59, 2095; V. Kalmeyer and R. B. Laughlin, Phys. Rev. B 39, 11 879; X. G. Wen, Frank Wilczek, and A. Zee, Phys. Rev. B 39,
11 413 (1989).
\bibitem{wen1991} X.-G. Wen, Phys. Rev. B 44, 2664 (1991).
\bibitem{senthil2000} T. Senthil and M. P. A. Fisher.  Phys. Rev. B 62, 7850 (2000).
\bibitem{fradkinshenker} E. Fradkin, S. Shenker, Phys. Rev. D 19, 3682-3697 (1979).
\bibitem{fradkin_raman} S. Papanikolaou, K. S. Raman, E. Fradkin, Phys. Rev. B 76, 224421 (2007).
\bibitem{HaldaneSastry}  F. D. M. Haldane, Phys. Rev. Lett. 60, 635 (1988). B. S. Shastry, Phys. Rev. Lett. 60, 639 (1988).
\bibitem{palee2005}S.-S. Lee and P. A. Lee, Phys. Rev. Lett. 95, 036403 (2005).
\bibitem{yamashita2010} M. Yamashita et al, Science 328, 1246 (2010). Y. Shimizu et al, Phys. Rev. Lett. 91, 107001 (2003). Y. Okamoto et al, Phys. Rev. Lett. 99, 137207 (2007).
\bibitem{hastings2010} M. B. Hastings et al., Phys. Rev. Lett. 104,
157201(2010).
\bibitem{gros1989} C. Gros, Ann. Phys. 189, 53 (1989).
\bibitem{frank2011crit} Yi Zhang, Tarun Grover, Ashvin Vishwanath, Phys. Rev. Lett. 107, 067202 (2011)
\bibitem{frank2011topo}Yi Zhang, Tarun Grover, Ashvin Vishwanath, Phys. Rev. B 84, 075128 (2011).
\bibitem{furukawa2007} S. Furukawa and G. Misguich, Phys. Rev. B 75, 214407 (2007).
\bibitem{haque2007} Masudul Haque, Oleksandr Zozulya, and Kareljan
Schoutens Phys. Rev. Lett. 98, 060401 (2007); O. S. Zozulya, M. Haque, K.
Schoutens, and E. H. Rezayi, Phys. Rev. B 76,
125310 (2007).
\bibitem{hamma2008} A. Hamma, W. Zhang, S. Haas, and D. A. Lidar, Phys. Rev. B 77, 155111 (2008).
\bibitem{isakov2011} Sergei V. Isakov, Matthew B. Hastings, Roger G. Melko,
arXiv:1102.1721.


\bibitem{thomale} D. F. Schroeter, E. Kapit, R. Thomale, and M. Greiter, Phys. Rev. Lett. 99, 097202 (2007).
\bibitem{ludwig1994} Andreas W. W. Ludwig, Matthew P. A. Fisher, R. Shankar, G.
Grinstein, Phys. Rev. B 50, 7526(1994).
\bibitem{tang2010} E. Tang, J. W. Mei, and X. G. Wen, arXiv:1012.2930.
\bibitem{sheng2011} Yi-Fei Wang, Zheng-Cheng Gu, Chang-De Gong, D. N. Sheng, arXiv:1103.1686;D. N. Sheng, Zheng-Cheng Gu, Kai Sun, L. Sheng, arXiv:1102.2658v1.
\bibitem{neupert2010}  T. Neupert, L. Santos, C. Chamon, and C. Mudry, arXiv:1012.4723.
\bibitem{sun2010} K. Sun, Z. C. Gu, H. Katsura, and S. Das Sarma, arXiv:1012.5864.
\bibitem{wen1990} X.-G. Wen, Int. J. Mod. Phys. B4, 239 (1990).
\bibitem{nayak} see e.g. Chetan Nayak, Steven H. Simon, Ady Stern, Michael Freedman, and Sankar Das Sarma, Rev. Mod. Phys. 80, 1083 (2008).
\bibitem{wen93} E. Keski-Vakkuri and Xiao-Gang Wen, Int. J. Mod. Phys. B7, 4227 (1993).
\bibitem{yellowbook}  P. Di Francesco, P. Mathieu, D. Senechal, {\em Conformal field theory,} Springer
1997.
\bibitem{kitaev_honey} A. Kitaev Ann. Phys. 321, 2 (2006).
\bibitem{Verlinde} Erik P. Verlinde, Nucl. Phys. B300:360, (1988).
\bibitem{Bais} F.A. Bais, J.C. Romers, arXiv:1108.0683v1.
\bibitem{dong2008} Shiying Dong, Eduardo Fradkin, Robert G. Leigh, Sean Nowling, JHEP 0805:016(2008).
\bibitem{frank2012} Yi Zhang, Tarun Grover, Ari Turner, Masaki Oshikawa, Ashvin Vishwanath, Phys. Rev. B 85, 235151 (2012).
\bibitem{wenplaq} X.-G. Wen, Phys. Rev. Lett. 90, 016803 (2003).
\bibitem{footnote_gsd} We note that there are exceptions to the equality between the number of degenerate ground states and the number of quasiparticle types. For example, the Wen-Plaquette model \cite{wenplaq} has a ground state degeneracy of two on an even $\times$ odd torus, even though it is $\mathbb{Z}_2$ topologically ordered with four quasiparticle types.

\bibitem{Cirac} J. I. Cirac, German Sierra,  Phys. Rev. B 81, 104431 (2010).
\bibitem{piflux} I. Affleck and J. B. Marston, Phys. Rev. B 37, R3774 (1988), J. B. Marston and I. Affleck, Phys. Rev. B 39, 11538 (1989).

\bibitem{wen2002} Xiao-Gang Wen, Phys. Rev. B 65, 165113 (2002).
\bibitem{li2011} Tao li, arXiv:1101.0193v1.
\bibitem{white1992} Steven R. White Phys. Rev. Lett. 69, 2863 (1992).
\bibitem{cincio}  Lukasz Cincio, Guifre Vidal, arXiv:1208.2623v1.


\end{thebibliography}
\end{document}